\documentclass[lettersize,journal]{IEEEtran}
\usepackage{amsmath,amsfonts}
\usepackage{algorithmic}
\usepackage{algorithm}
\usepackage{array}
\usepackage[caption=false,font=normalsize,labelfont=sf,textfont=sf]{subfig}
\usepackage{textcomp}
\usepackage{stfloats}
\usepackage{url}
\usepackage{verbatim}
\usepackage{graphicx}
\usepackage{cite}
\usepackage{multirow}
\usepackage{makecell}

\begin{document}

\title{Hazard Resistance-Based Spatiotemporal \\ Risk Analysis for Distribution Network Outages During Hurricanes}
\author{Luo Xu,~\IEEEmembership{Member,~IEEE}, Ning Lin, Dazhi Xi, Kairui Feng, and H. Vincent Poor,~\IEEEmembership{Life Fellow,~IEEE}
\thanks{This work was supported by US National Science Foundation grant numbers 1652448 and 2103754 (as part of the Megalopolitan Coastal Transformation Hub) and Princeton University Innovation Fund. \textit{(Corresponding author:  Luo Xu.)}}
\thanks{L. Xu, N. Lin, D. Xi, and K. Feng are with the Department of Civil and Environmental Engineering, Princeton University, Princeton, NJ, 08544, USA. L. Xu and N. Lin are also with Center for Policy Research on Energy and the Environment, Princeton University, Princeton, NJ, 08544, USA (e-mail: luoxu@princeton.edu, nlin@princeton.edu).}
\thanks{H. V. Poor is with the Department of Electrical and Computer Engineering, Princeton University, Princeton, NJ, 08544, USA. (e-mail: poor@princeton.edu).
}}
\markboth{Journal of \LaTeX\ Class Files,~Vol.~14, No.~8, August~2021}%
{Shell \MakeLowercase{\textit{et al.}}: A Sample Article Using IEEEtran.cls for IEEE Journals}


\maketitle

\begin{abstract}
Blackouts in recent decades show an increasing prevalence of power outages due to extreme weather events such as hurricanes. Precisely assessing the spatiotemporal outages in distribution networks, the most vulnerable part of power systems, is critical to enhance power system resilience. The Sequential Monte Carlo (SMC) simulation method is widely used for spatiotemporal risk analysis of power systems during extreme weather hazards. However, it is found here that the SMC method can lead to large errors by directly applying the fragility function or failure probability of system components in time-sequential analysis, particularly overestimating damages under evolving hazards with high-frequency sampling. To address this issue, a novel hazard resistance-based spatiotemporal risk analysis (HRSRA) method is proposed. This method converts the time-varying failure probability of a component into a hazard resistance as a time-invariant value during the simulation of evolving hazards. The proposed HRSRA provides an adaptive framework for incorporating high-spatiotemporal-resolution meteorology models into power outage simulations. By leveraging the geographic information system data of the power system and a physics-based hurricane wind field model, the superiority of the proposed method is validated using real-world time-series power outage data from Puerto Rico during Hurricane Fiona 2022.
\end{abstract}


\section{Introduction}
\IEEEPARstart{A}{s} one of the most critical infrastructures of modern societies, power systems have been designed to supply electricity under normal and abnormal circumstances, e.g., system contingencies \cite{ref1,ref2}. However, recent years have witnessed increasing numbers of catastrophic outages in power systems under extreme weather events \cite{ref0}. For instance, both Hurricane Fiona in 2022 and Hurricane Maria in 2017 brought down the Puerto Rico power grid, resulting in complete blackouts \cite{ref3,ref4}. Reports indicate that over 80\% of major power outage events in the U.S. between 2000 and 2021 were weather-related. According to the North American Electric Reliability Corporation \cite{ref5}, the number of power outages caused by extreme weather events is on the rise; annual weather-related power outages have increased by about 78\% in this decade compared to the last decade \cite{ref6}.

Compared to transmission networks, distribution networks, which have lower infrastructure resistance, are more susceptible to extreme weather events such as hurricanes. The Puerto Rico power utility reported that more than half of its distribution feeders were damaged during Hurricane Fiona in September 2022 \cite{ref7}. Analyzing the spatiotemporal power outages in distribution networks during extreme weather events is critical for fault detection and localization \cite{ref8}. Accurately forecasting outages in distribution networks is also beneficial for capturing supply-demand imbalances in power system dispatch and emergency control, ensuring system stability. Moreover, by understanding these outages’ spatiotemporal patterns, system operators can more accurately predict future outages under intensifying extreme weather events in a changing climate. This understanding will further assist power utilities in designing climate-resilient networks \cite{ref9}.

To investigate weather-associated spatiotemporal outages of power systems, the Sequential Monte Carlo (SMC) simulation method has been extensively adopted in existing studies \cite{ref10,ref11,ref12,ref13,ref14,ref15,ref16}. Compared to the Markov method, which is more suitable for small-scale systems with lower computational burden in analyzing stationary random processes such as common cause failures in power system reliability analysis \cite{ref17,ref18}, the SMC simulation method serves as a powerful tool for risk assessment, capable of handling complex systems with numerous components. The SMC method determines the time-varying failure probability of each component at every time interval based on the fragility as a function of weather intensity \cite{ref10}. Subsequently, the operational status of the component is sampled by generating a uniform random number that is compared to the time-varying failure probability.

However, it is noteworthy that most existing infrastructure fragility functions are primarily hazard intensity-dependent without consideration of the specific time duration \cite{ref19,ref20}. Applied to model power outages during persistent, temporally evolving hazards such as hurricanes, the SMC simulation method becomes sensitive to the sampling frequency. When integrated with high-resolution spatiotemporal hazard models, the SMC method generates simulation results with what can be termed the \textbf{‘failure probability curse’}: a contradiction between sampling frequency and simulation precision. High-frequency sampling can lead to an overestimation of power outages due to the temporal accumulation of the time-varying failure probability for each component. Even if the time-varying failure probability is scaled by factoring in the number of time intervals \cite{ref11,ref21}, the issue of sampling frequency remains unsolved since time intervals are determined by the arbitrarily chosen simulation duration. Moreover, due to the lack of comparison with real-world spatiotemporal power outage cases and the absence of applying high-resolution meteorology models for hazard simulations, this issue has not been prominently highlighted. 

To address this issue, we develop a high-precision spatiotemporal outage simulation method that adapts to the hazard simulation resolution. We propose a novel spatiotemporal risk analysis framework that accounts for the uncertainty of time-invariant hazard resistance. Unlike the SMC simulation method that applies the failure probability in time series analysis, the proposed method considers the hazard resistance to be an inherent characteristic of a component and time-invariant throughout the temporally evolving hazard simulation. The hazard resistance is derived by converting it from the failure probability. 

Based on the proposed risk assessment method, this paper performs a real-world case study that is traced back to the power outage in Puerto Rico during Hurricane Fiona in September 2022. The contributions of this paper are summarized as follows.

(1) This paper provides an interdisciplinary study that intersects models and datasets from power system and meteorology fields. The high-resolution geographic information system (GIS) data of the Puerto Rico distribution network with a feeder-level load profile is collected and introduced in this paper. For the hazard simulation, a physics-based hurricane wind model for generating the spatiotemporal wind field is incorporated. 

(2) A novel hazard resistance-based spatiotemporal risk analysis (HRSRA) method for time-series distribution network outages is introduced. By converting the failure probability into the hazard resistance of each component, our proposed risk analysis framework addresses the challenges posed by the failure probability curse and is adaptive to varying resolutions of hazard simulation.

(3) We validate the superiority of the spatiotemporal simulation results from the proposed HRSRA against the SMC simulation method using observed outage data from Puerto Rico, recorded by the local power utility during Hurricane Fiona.

This paper is organized as follows. Following the introduction, Section II introduces the proposed methods and datasets used for quantifying the spatiotemporal distribution network outages during hurricanes. In Section III, the proposed methods are validated by the real-world case study. Section IV concludes this paper.

\section{Methods and Datasets}
\begin{figure}[!t]
\centering
\includegraphics[width=3.5in]{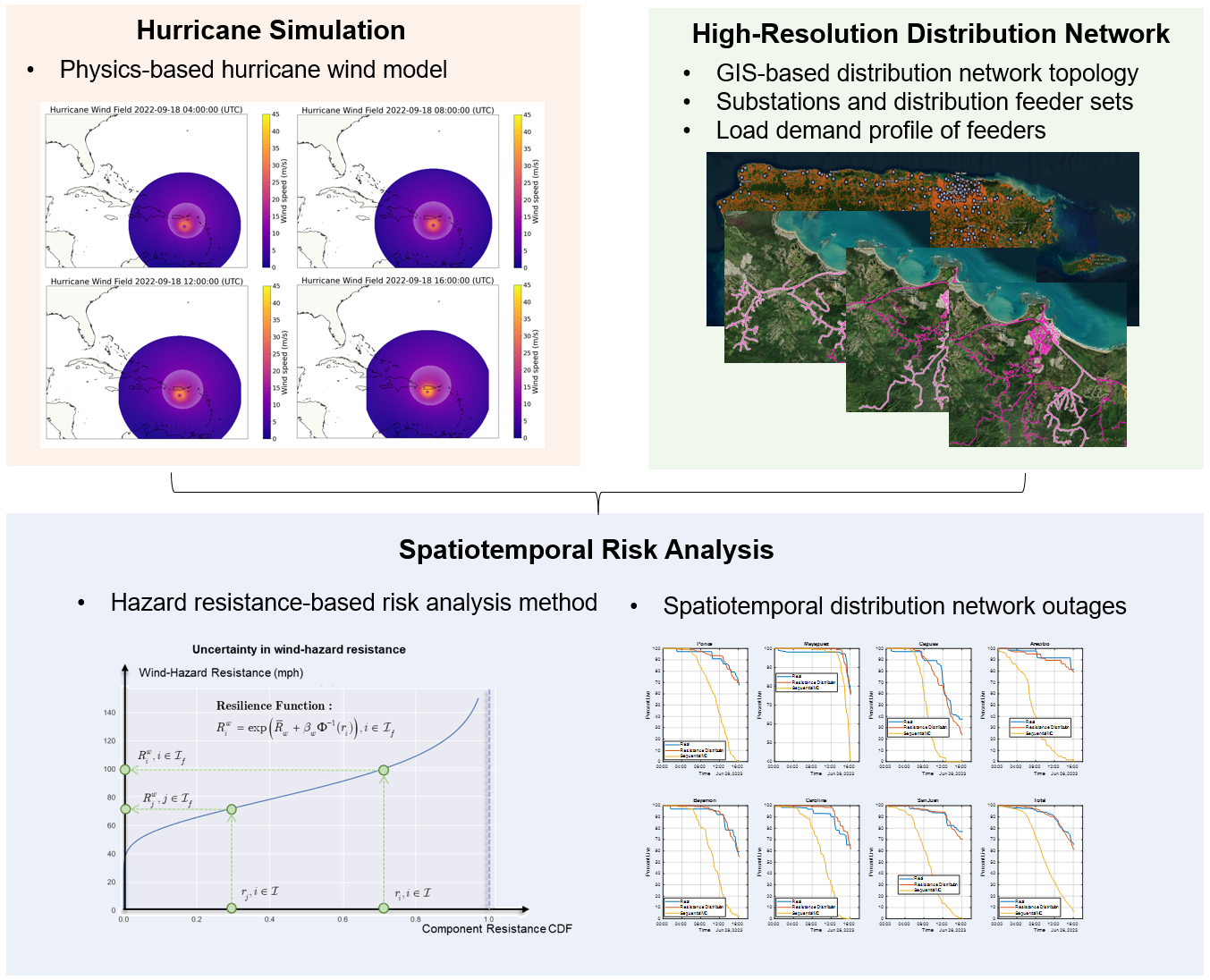}
\caption{Overview of the proposed framework for quantifying spatiotemporal power outage in distribution networks during hurricane events.}
\label{Fig_1}
\end{figure}

An overview of the proposed methods and the datasets used is shown in Fig. \ref{Fig_1}. We use a physics-based spatiotemporal tropical cyclone wind field model to simulate the hurricane hazard. The hazard simulation is applied to a high-resolution distribution network model of Puerto Rico using geographic information system (GIS) data and feeder-level power flow datasets. Based on the meteorology and power system models, we develop a hazard resistance-based risk analysis framework that is used to quantify the spatiotemporal power outage during non-stationary extreme weather events. Instead of sampling the time-varying failure probability, we establish a time-invariant resistance distribution for power system infrastructure and avoid the failure probability curse due to high-frequency sampling. Under high-resolution time-varying hazard simulations, this risk analysis framework enables more accurate estimation of distribution network outages.

\subsection{Puerto Rico Distribution Network}

\subsubsection{High-Resolution GIS Distribution Network}  

High-resolution GIS data of the Puerto Rico distribution network infrastructure was obtained from the Puerto Rico Electric Power Authority (PREPA) through Puerto Rico Innovation and Technology Service \cite{ref22}. The Puerto Rico distribution network is characterized by four voltage levels (4 kV, 7 kV, 8 kV and 13 kV). This high-resolution GIS data includes the topology of the Puerto Rico distribution network, coordinates of substations, and coordinates of distribution poles.

Load demand profile datasets of all 296 substations with 936 operational distribution feeders updated in 2022 were sourced from PREPA and LUMA Energy, the local system operator and power utility of the Puerto Rico power grid serving a total population of 3.2 million \cite{ref7}. The total peak load demand of Puerto Rico is 2751 MW.

\begin{figure}[H]
\centering
\includegraphics[width=3.5in]{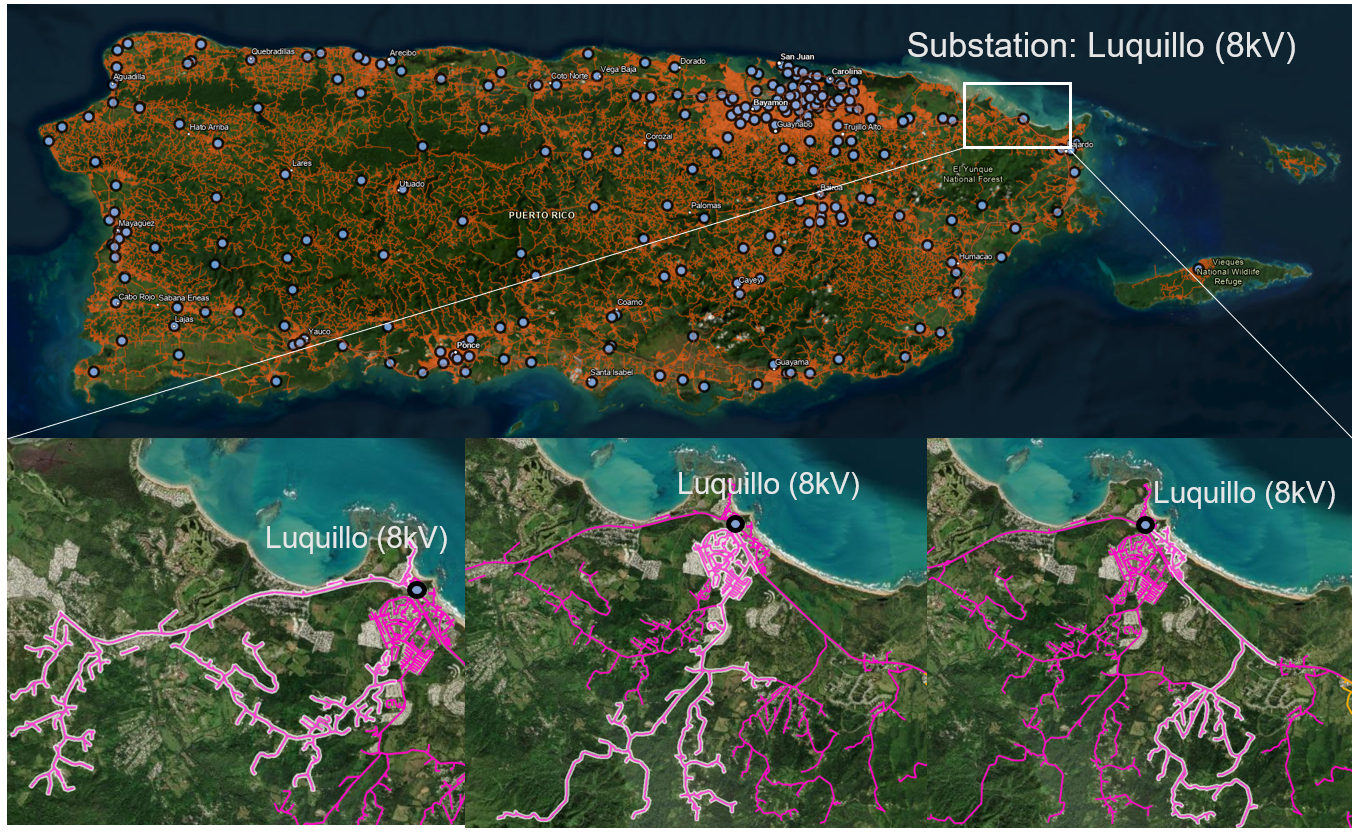}
\caption{High-resolution geographic information system data of the Puerto Rico distribution network: An illustration of the Luquillo Substation at 8kV with three distribution feeders, each highlighted in white at the bottom.}
\label{Fig_2}
\end{figure}

Fig. \ref{Fig_2} depicts the example of an 8kV substation named Luquillo, located on the northeast coast of Puerto Rico. The figure displays GIS information for its three distribution feeders, with each feeder highlighted in white. The detailed feeder-level GIS information of Puerto Rico distribution networks enables the high-resolution spatiotemporal comprehensive risk analysis of the networks’ performance during hurricane events.

\subsubsection{Power Outage Data}  
\begin{figure}[H]
\centering
\subfloat[]{\includegraphics[width=3.4in]{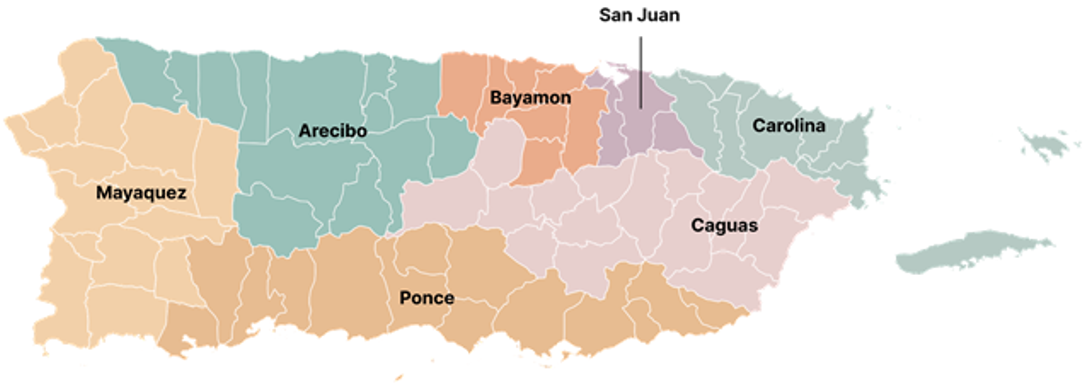}%
\label{Fig_3a}}
\hfil
\subfloat[]{\includegraphics[width=3.4in]{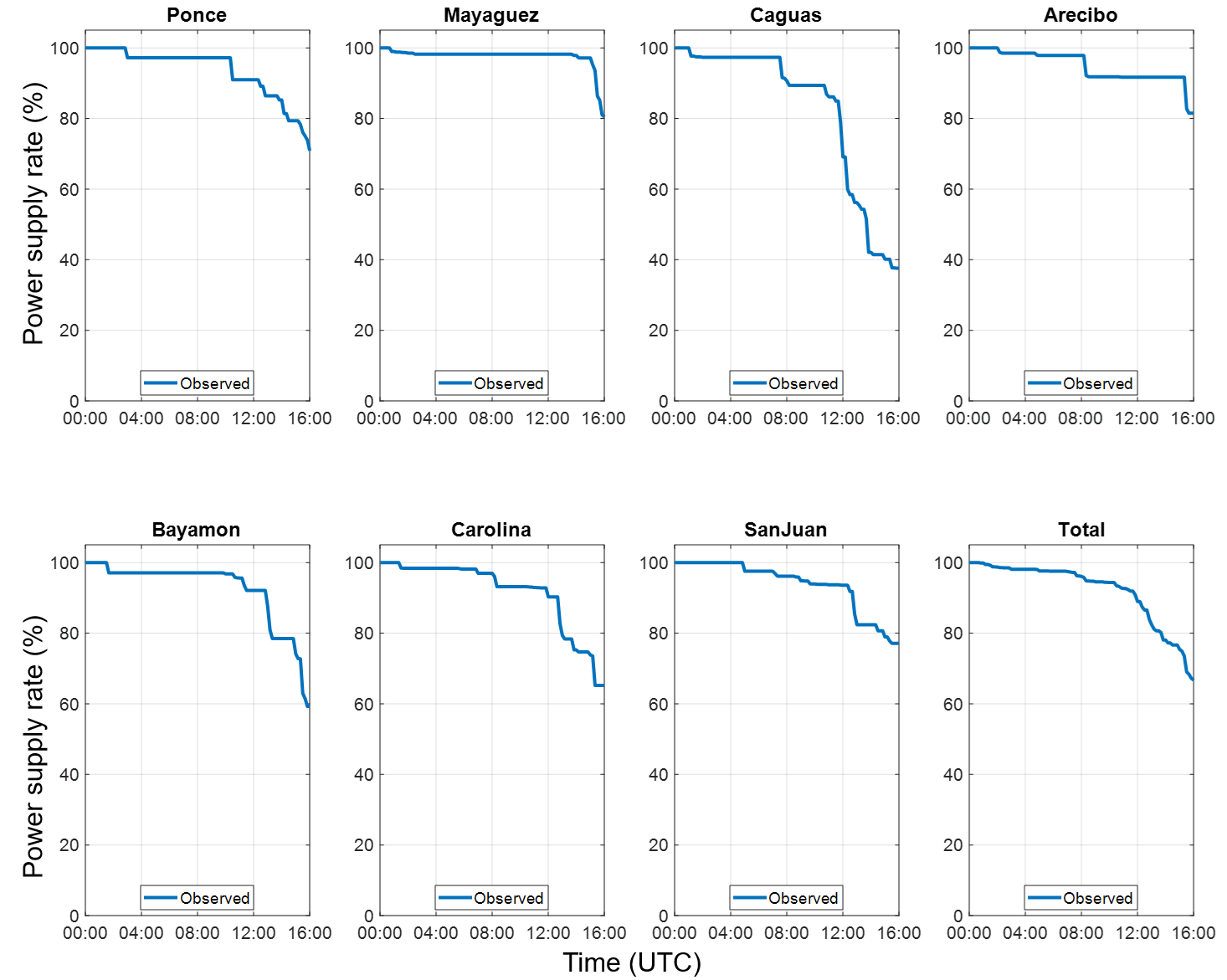}%
\label{Fig_3b}}
\caption{Seven regions of the Puerto Rico power grid and power supply rate. (a) Puerto Rico power grid operational divisions. (b) Regional and total power outage data.}
\label{Fig_3}
\end{figure}

Spatiotemporal customer outage data of the Puerto Rico power grid during Hurricane Fiona on September 18, 2022, was recorded every 10 minutes by LUMA Energy \cite{ref23}. The power outage datasets were categorized into seven distinct regions: Ponce, Mayaguez, Arecibo, Bayamon, Carolina, Caguas, and San Juan, as shown in Fig. \ref{Fig_3}(a), which are consistent with PREPA’s operational divisions.

\begin{figure}[H]
\centering
\includegraphics[width=3.5in]{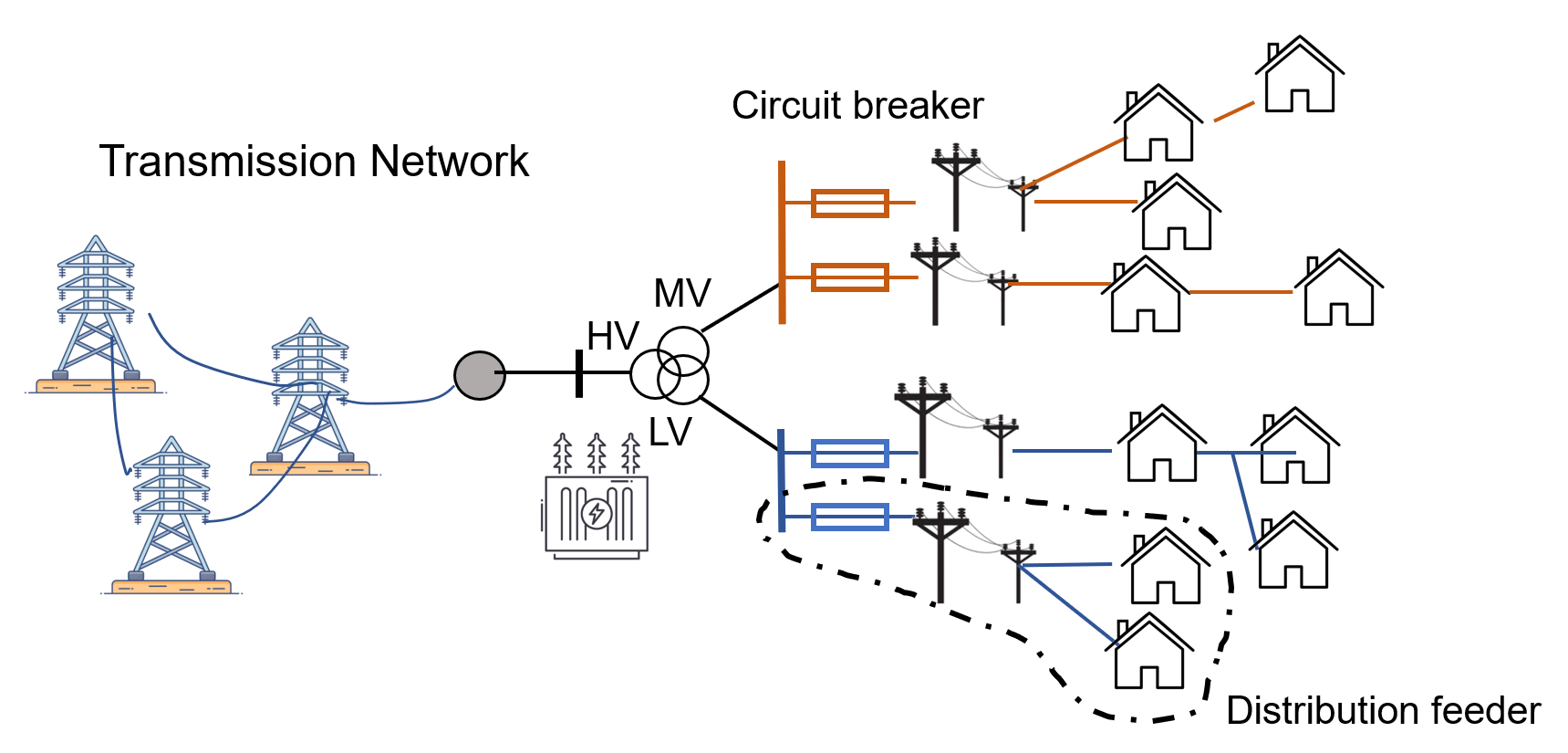}
\caption{Diagram of a distribution network.}
\label{Fig_4}
\end{figure}

Although a catastrophic blackout (from over 50\% supply rate directly to 0\%) happened in the Puerto Rico power grid at approximately 18:00 UTC on September 18, 2022, during Hurricane Fiona, we note that the power outage in the early stage of this event is due to distribution network failures rather than transmission network failures. Therefore, to focus on the behavior of the distribution network, we chose the spatiotemporal power outage data in the time interval between 0:00 UTC (before the occurrence of damage) and 16:00 UTC, September 18, 2022, as a basis to validate our proposed method. The power outage data related to corresponding regions is shown in Fig. \ref{Fig_3}(b).
As most distribution networks of power grids operate on radial topologies, as illustrated in Fig. \ref{Fig_4}, each substation consists of several distribution feeders that deliver electricity to end users. Damage to distribution poles or lines due to extreme wind events within a distribution feeder will trigger the overcurrent relay protection of the circuit breaker or the fuse cut-out due to short-circuit conditions, resulting in the whole feeder losing power \cite{ref24,ref25}. Compared to outages at the substation level, outages at the distribution feeder level are more representative of distribution network faults and commonly simulated in the existing research \cite{ref14,ref26}. Therefore, simulated power outages based on the spatiotemporal risk analysis method in this paper are analyzed at the feeder level to compare with the real-world outage data.

\subsection{Tropical Cyclone Wind Field Simulation}
There are several methods that are commonly employed to simulate tropical cyclone (TC) winds. The full-physics numerical models, such as the Weather Research and Forecast (WRF) model \cite{ref27}, calculate TC winds by solving the primitive equations that govern the atmospheric dynamic and thermodynamic processes. While these methods are recognized for their accuracy, they might not provide the most accurate estimation for hindcasting of historical storms, and their computational intensity makes them less suitable for high-resolution spatiotemporal hazard assessment tasks. To efficiently evaluate the wind hazards in power system risk analysis, TC wind profile models (e.g., the Holland model \cite{ref28}) driven by observed TC characteristics are used to simulate symmetric wind fields of TCs in state-of-the-art power systems research \cite{ref11,ref12,ref13,ref16,ref29}. However, given the combined effects of a hurricane’s rotation and translation, its actual wind field exhibits significant asymmetry. Using symmetric models can lead to large inaccuracies in high-resolution spatiotemporal power outage assessment.

\begin{figure*}[!t]
\centering
\includegraphics[width=7in]{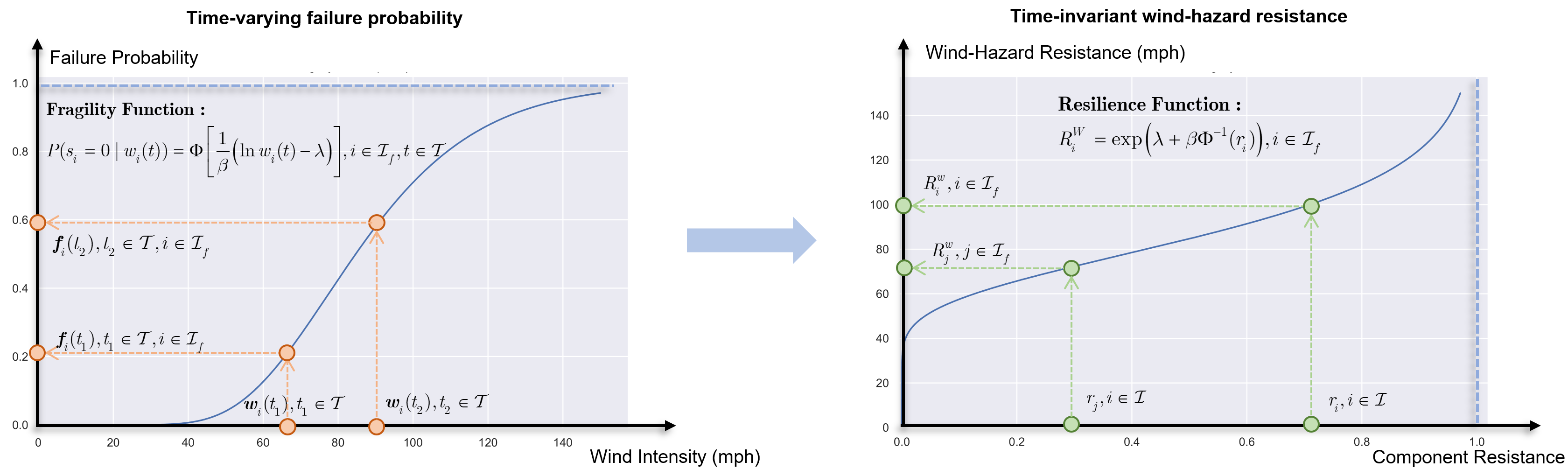}
\caption{Comparison of uncertainties considered in the Sequential Monte Carlo (SMC) method (Left) and in the proposed HRSRA (Right). Both methods are based on the same fragility curve. SMC samples the system state based on a fragility curve with uncertainty in the time-varying failure probability. We redefine the uncertainty into the time-invariant hazard resistance distribution of all components.}
\label{Fig_5}
\end{figure*}

In this study, we employ a physics-based wind profile model \cite{ref30} that considers both the TC inner core and outer radii dynamics to simulate the wind profile. Then by combining the TC wind profile with the environmental background wind from TC translation speed according to \cite{ref31}, the high-resolution asymmetric wind fields for a hurricane can be generated efficiently. The accuracy of this physics-based model has also been validated and utilized for simulating TC-related hazards such as wind, rainfall, and storm surges in risk analysis \cite{ref32,ref33}. To simulate the wind field of Hurricane Fiona, we use historical TC track data (Hurricane Fiona in 2022) obtained from the International Best Track Archive for Climate Stewardship (IBTrACS) at the National Center for Environmental Information of the National Oceanic and Atmospheric Administration (NOAA) \cite{ref34}. This dataset provides a time-series of TC locations, maximum sustained winds, and radius of maximum winds.

The spatiotemporal TC model generates a 1-minute sustained wind field. However, due to the compounded effects of wind turbulence and sustained wind, the resulting impact on infrastructure is substantially magnified. Consequently, infrastructure fragility analysis is widely conducted under the conditions of 3-second gust wind \cite{ref35,ref36}. Therefore, we convert the sustained wind into a gust wind field by multiplying it by a 3-second gust factor. Here, the 3-second gust wind factor $G_\tau$ for 1-minute sustained wind is chosen to be 1.49 according to \cite{ref35}.

\subsection{Hazard-Resistance Spatiotemporal Risk Analysis Method}

Instead of using the SMC method, we propose a novel HRSRA method for time-series of distribution network outages during the wind intensity evolution process. The uncertainty considered in the SMC method comes from the time-varying failure probabilities of each component. However, the proposed HRSRA method considers that the uncertainty of power outage simulation lies in the probability distribution of the distribution feeders’ hazard (wind) resistances. 

For given distribution feeder sets $\mathcal{I}_f$ in the distribution network $\mathcal{G}$, fragility functions are used to represent the relationship between component failure probability and hazard intensity. In particular, the material strength properties of power system components against natural hazards are widely recognized to have a lognormal distribution \cite{ref10,ref19}. Therefore, the fragility curve for the distribution feeder is characterized as the lognormal cumulative distribution function (CDF), and the failure probability of a distribution feeder under a certain hazard intensity is determined by
\begin{equation}
\label{CDF}
P(s = 0|w) = \Phi \left[ {\frac{1}{\beta }\left( {\ln w - \lambda } \right)} \right]
\end{equation}

\noindent where $s$ refers to the damage state of the distribution feeder, and here $s=0$ indicates the component is damaged; $\beta$ is the standard deviation of the logarithm of engineering stress parameter where a component reaches the threshold of the damage state $s=0$; $\lambda$ is the mean of the logarithm of the engineering stress parameter; $\Phi$ denotes the CDF of normal distribution.

As shown in Fig. \ref{Fig_5}, in the SMC method, the time-varying failure probability of component $i \in \mathcal{I}_f$, that is, ${f_i}(t) = P({s_i} = 0|{w_i}(t))$ is calculated based on (\ref{CDF}) combined with the current wind intensity at every time step $t \in \mathcal{T}$. Then, a uniformly distributed random number ${r_i}(t) \sim U(0,1)$ is generated at every time step $t \in \mathcal{T}$ for sampling the failure status as follows:
\begin{equation}
\label{SMC}
\begin{array}{l}
s_i^{SMC}(t) = \left\{ \begin{array}{l}
0,{\rm{ }}{r_i}(t) \le {f_i}(t)  \text{ or }  s_i^{SMC}(t - 1) = 0\\
1,{\rm{ else }}
\end{array} \right.\\
i \in {{\cal I}_f},t \in {\cal T}
\end{array}
\end{equation}

\noindent where $s_i^{SMC}(t)$ is 0 if the distribution feeder experiences an outage at time step $t \in \mathcal{T}$ or earlier, and 1 otherwise.

It can be observed from equation (\ref{SMC}) that the status of the component is examined at every time step during this time series analysis. Thus, a higher sampling frequency results in a higher failure probability when applying the SMC method. Therefore, there is a contradiction between the high-resolution hazard simulation and the accuracy of the spatiotemporal risk analysis.

To avoid this failure probability curse, the proposed HRSRA converts the failure probability to the hazard resistance distribution. As illustrated in Fig. \ref{Fig_5}, we define the hazard resistance for distribution feeder $i \in \mathcal{I}_f$ to be $R_i^W$, indicating failure $s_i = 0$ when the wind intensity exceeds $R_i^W$. This hazard resistance is time-invariant but varies over the distribution feeders, according to the fragility function. Given that the probability of a component’s hazard resistance being less than a hazard intensity equates to its failure probability under the hazard intensity, that is, the CDF of $R_i^W$ satisfying $F_{R_i^W}(w) = P(R_i^W \le w) = P({s_i} = 0|w)$, one can generate samples of   using the inverse transformation method and Equation (\ref{CDF}), as
\begin{equation}
\label{HRSRA}
\begin{array}{l}
R_i^W = F_{R_i^W}^{ - 1}({r_i})\\
 \quad  \quad = \exp \left( {\lambda  + \beta  \cdot {\Phi ^{ - 1}}({r_i})} \right),i \in {{\cal I}_f}
\end{array}
\end{equation}

\noindent where ${\Phi ^{ - 1}}( \cdot )$ is the inverse of the standard normal CDF and ${r_i} \sim U(0,1)$ is a sample from the standard uniform distribution. By using the above hazard resistance function, the time-varying sample space of failure probability is mapped into a time-invariant sample space of hazard resistance. In every outage simulation using the proposed HRSRA method, a distribution feeder is assigned a time-invariant hazard resistance that remains constant throughout the time-series analysis.

\begin{algorithm}
\caption{Distribution Network Outage Simulation with Hazard-Resistance Uncertainty}
\begin{algorithmic}[1]
\STATE \textbf{Data Input:} Number of simulations $N$, distribution network $\mathcal{G}$, sets of distribution feeder $\mathcal{I}_f$, simulation time period $\mathcal{T}$, hurricane track data;
\STATE Initialize counter, distribution feeder status $s_i(0)=1, i \in \mathcal{I}$, system failure status $\boldsymbol{p}_{\text{fail},j} = \boldsymbol{0}$;
\FOR{$j = 1$ \TO $N$}
    \STATE Initialize system status and system loss $\boldsymbol{L}_{sys}$;
    \STATE Generate a random number $r_i \sim U(0,1)$ for the $i$-th feeder;
    \STATE Calculate hazard resistance of each distribution feeder by $R_i^w = \exp({{{\bar R}_w} + {\beta _w}{\Phi ^{ - 1}}(r_i)}), i \in \mathcal{I}_f$;
    \FOR{$t \in \mathcal{T}$}
        \STATE Generate boundary-layer spatiotemporal wind field of hurricane ${{\cal W}^{\phi  \times \psi  \times t}}:w(\phi ,\psi ,t),t \in {\cal T}$;
        \STATE Transfer the sustained wind field to 3-second gust wind field ${w_G}(\phi ,\psi ,t) = {G_\tau } \cdot w(\phi ,\psi ,t)$;
        \FOR{$i \in \mathcal{I}_f$}
            \STATE Calculate the 3-second gust wind speed \\
            ${w_G}({\phi _i},{\psi _i},t)$ at location $(\phi_i, \psi_i)$ at time $t$;
            \IF{$R_i^W < {w_G}({\phi _i},{\psi _i},t)$  or $s_i(t-1)=0$}
                \STATE The $i$-th feeder fails, record $s_i(t) = 0$;
                \STATE Calculate system loss ${L_{sys}}(t)= {L_{sys}}(t) + {L_i}(t)$;
            \ELSE
                \STATE Record $s_i(t) = 1$, $L_{sys}(t) = {L_{sys}}(t)$;
            \ENDIF
        \ENDFOR
        \STATE Calculate system failure status at time $t$ : ${p_{{\rm{fail,}}j}}(t) = {L_{sys}}(t)/\sum\nolimits_{i \in {{\cal I}_f}} {{L_i}(t)} $;
    \ENDFOR
\ENDFOR
\RETURN $ \boldsymbol{p}_{\text{fail}} = [\boldsymbol{p}_{\text{fail},1}, \ldots, \boldsymbol{p}_{\text{fail},N}]$.
\end{algorithmic}
\end{algorithm}

Algorithm 1 summarizes the procedures of the proposed HRSRA method for the distribution network outage simulation. For a given distribution network $\mathcal{G}$ with a set of distribution feeders $\mathcal{I}_f$. In this Algorithm, we execute the hazard resistance sampling for all distribution feeders prior to the time-series simulation. The hazard resistance of a distribution feeder is time-invariant during the time-series analysis. At simulation time $t \in \mathcal{T}$, the wind speed for distribution $i$, that is, $w({\phi _i},{\psi _i},t)$ is calculated at location $({\phi _i},{\psi _i})$ based the generated hurricane wind field and the GIS datasets of distribution feeders. Then, we assign the following binary status to indicate the operation condition of the distribution feeder $i$  during the evolving hazard, which is described as the following equation:
\begin{equation}
\label{HRSRA_sample}
{s_i}(t) = {s_i}(t - 1) \cdot {I_{{w_G}({\phi _i},{\psi _i},t) > R_i^W}},i \in {{\cal I}_f},t \in {\mathcal{T}}
\end{equation}

\noindent where $s_i (t)$ is a binary variable with values 1 or 0 denoting the functional or the malfunctional status of the distribution feeder, respectively; $I$ denotes an indicator function that is 0 if current gust wind ${w_G}({\phi _i},{\psi _i},t) > R_i^W$ , and 1 otherwise. If the distribution feeder malfunctions, the load demand $L_i (t)$ will be included in the total system loss $L_{sys}(t)$ at simulation time $t \in \mathcal{T}$.

\section{Validation}
In this section, the distribution network outages of Puerto Rico during Hurricane Fiona in 2022 are retraced. The wind field simulation of Hurricane Fiona is generated using the wind profile model driven by observed track. The proposed spatiotemporal risk analysis framework for distribution network outage is validated using the actual outage records obtained during this catastrophic event. The results yielded by this proposed method are then compared with those generated by the SMC method, which serves to further validate its effectiveness and accuracy.

\begin{figure}[H]
\centering
\includegraphics[width=3.5in]{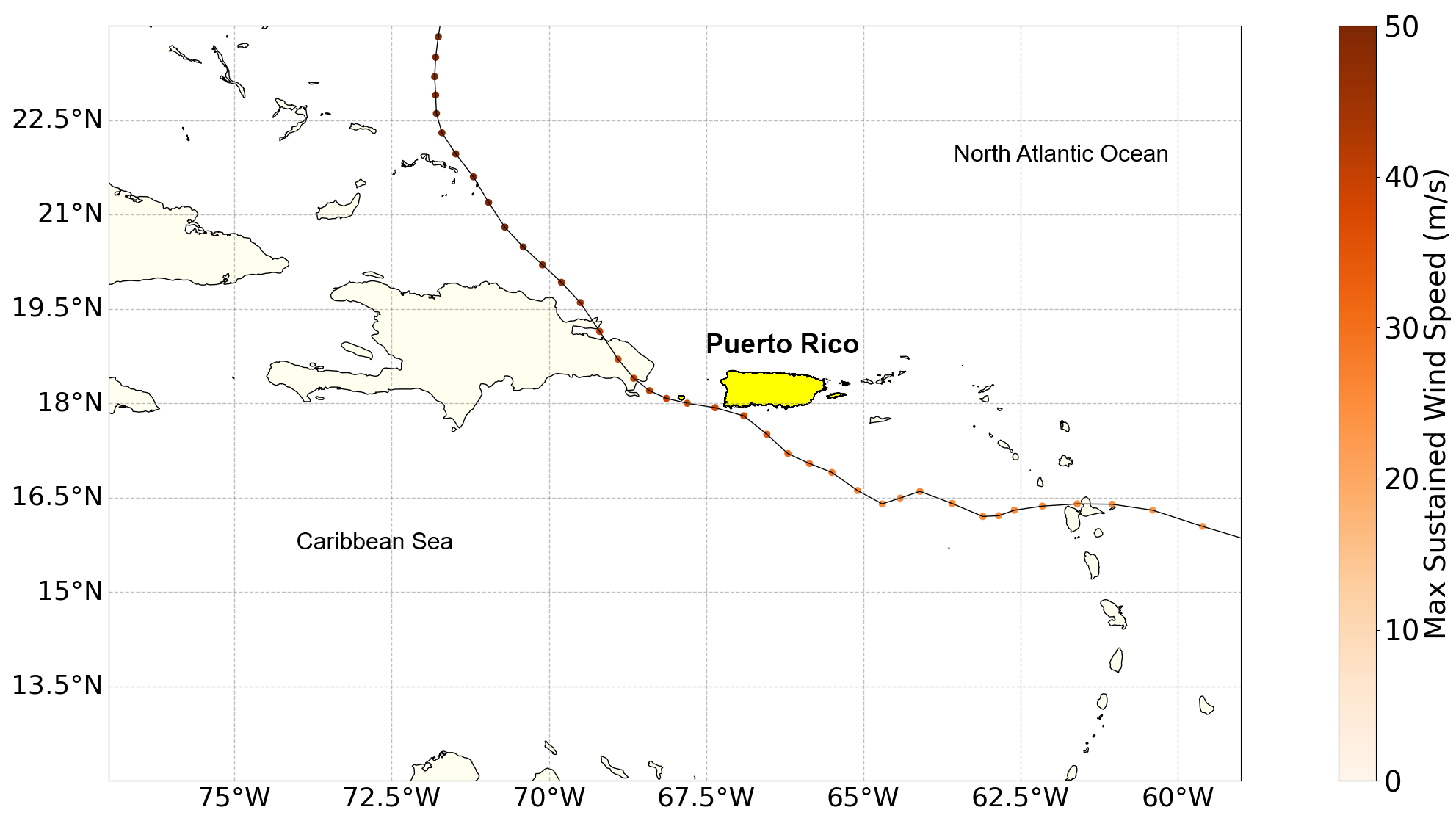}
\caption{Track and max sustained wind speed (m/s) of Hurricane Fiona in 2022.}
\label{Fig_6}
\end{figure}

\subsection{Hazard Simulation}
\begin{figure}[H]
\centering
\includegraphics[width=3.5in]{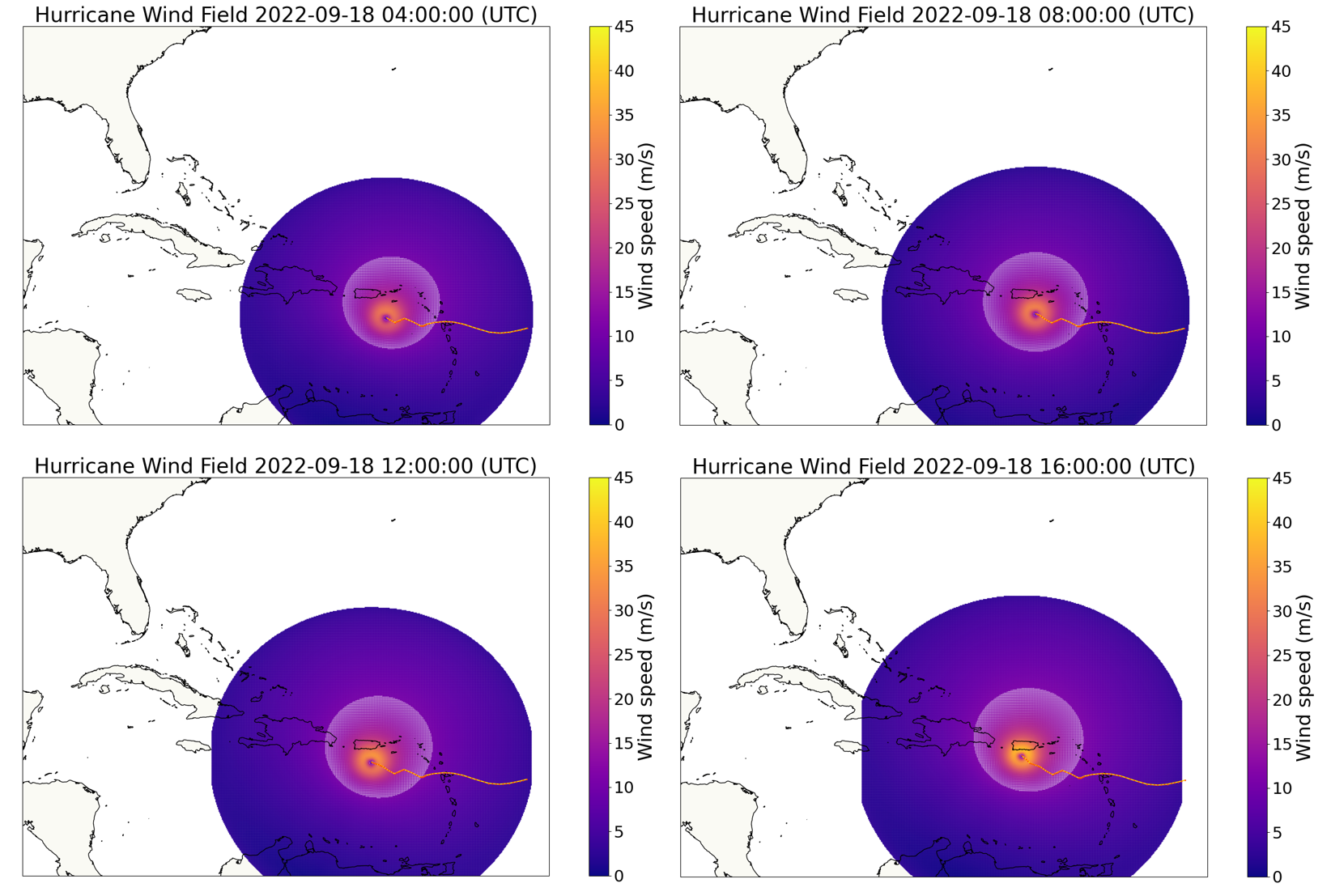}
\caption{Spatiotemporal wind field simulation of Hurricane Fiona in 2022.}
\label{Fig_7}
\end{figure}

On September 18, 2022, Hurricane Fiona made landfall in southwest Puerto Rico and caused a catastrophic blackout that affected more than three million residents. Fig. \ref{Fig_6} shows the track and max sustained wind speed of Hurricane Fiona in 2022, as provided by the IBTrACS dataset \cite{ref34}. The max sustained wind speed reached over 40 m/s. To simulate the high-resolution spatiotemporal distribution outage, the IBTrACS track data is interpolated into 10-minute intervals, and the wind field is generated for every 10 minutes during the hurricane event. The generated spatiotemporal wind fields of Hurricane Fiona at selected times are shown in Fig. \ref{Fig_7}. 

It can be observed from Fig. 7 that the boundary-layer wind speed within the Puerto Rico region demonstrated a gradual increase from 04:00 UTC to 16:00 UTC to over 40 m/s as the hurricane approaches and intensifies. The asymmetric wind fields demonstrate a heightened wind hazard on the right side of the storm’s track due to the counterclockwise rotation of the storm and the environmental wind aligned with its northwestward movement.

\begin{figure}[H]
\centering
\includegraphics[width=3.5in]{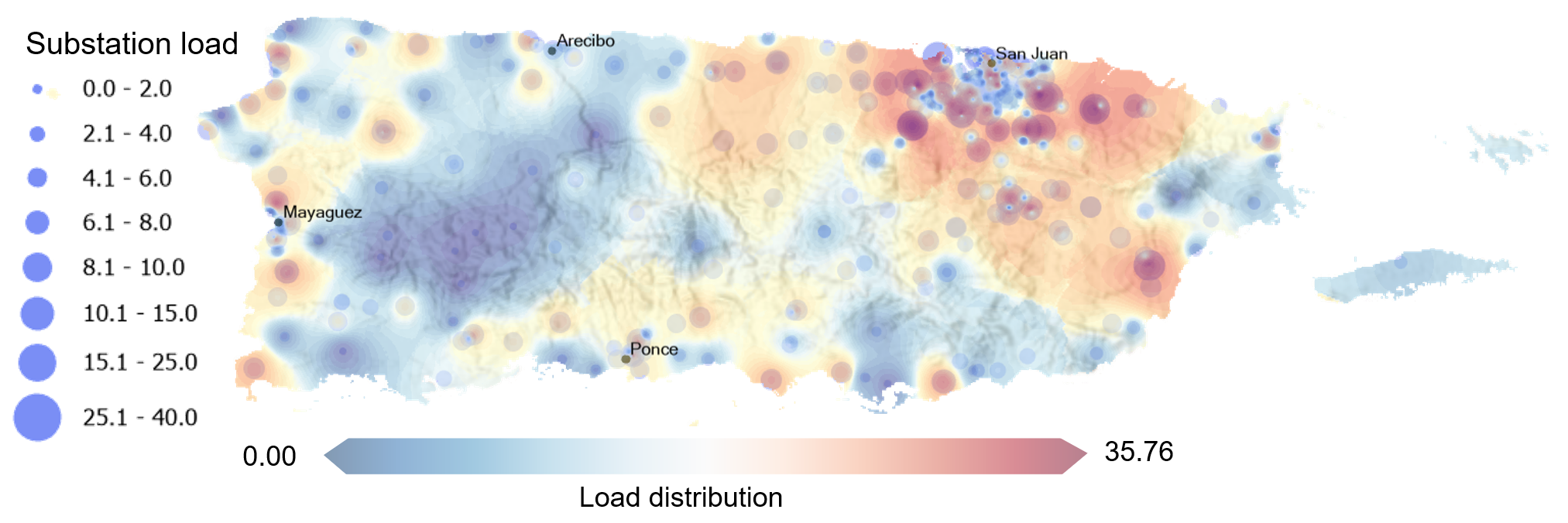}
\caption{Load demand distribution of the Puerto Rico power grid.}
\label{Fig_8}
\end{figure}

\subsection{Distribution Network Outage Simulation and Comparison}

The GIS information and distribution feeder datasets of the Puerto Rico distribution network are introduced in Section II. A. By aggregating the demand of distribution feeders under the same substation, the load distribution across the Puerto Rico power grid at the substation level is visualized in Fig. \ref{Fig_8}. This figure underscores the significant spatial variations in the grid’s demand profile. The real-world spatiotemporal outage datasets were categorized into seven regions by the local power utility, as shown in Fig. \ref{Fig_3}. Considering the distinct regional variations in both load and network structure, we apply region-specific lognormal fragility functions for the distribution network outage simulation. The fragility functions are calibrated based on the simulated wind fields and power outage rates from the observation during Hurricane Fiona. To show the superiority of the proposed method, the same fragility functions are applied to both the HRSRA and the SMC methods for spatiotemporal risk analysis. The detailed parameters of the fragility functions are provided in Appendix A.

\begin{figure}[H]
\centering
\includegraphics[width=3.5in]{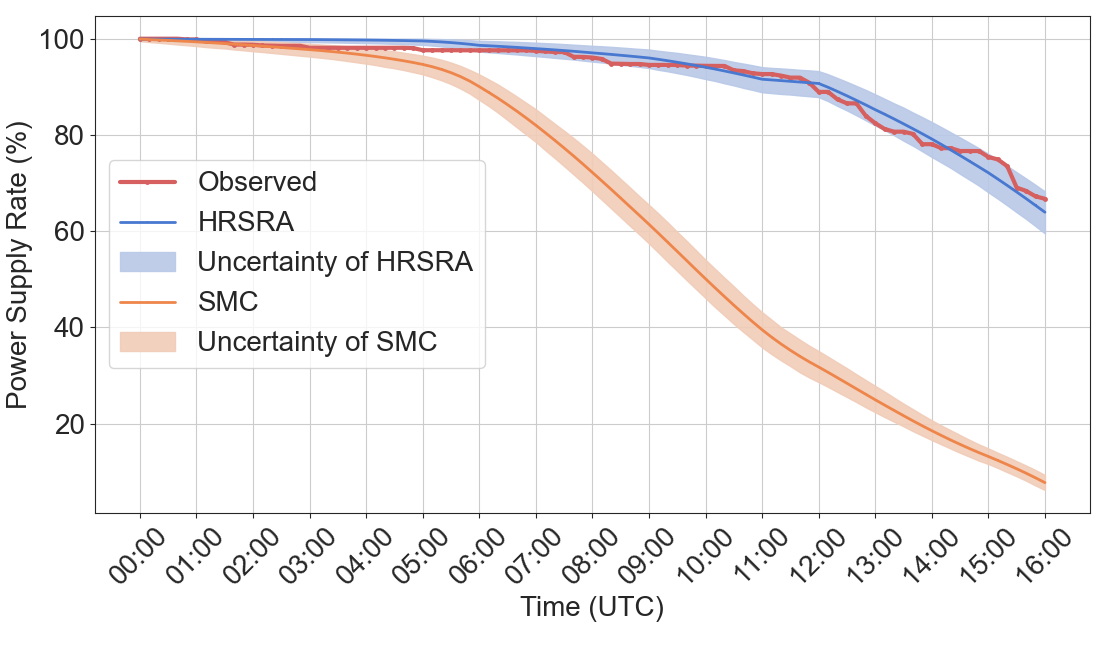}
\caption{Observed and simulations of total distribution network outages during Hurricane Fiona on September 18, 2022.}
\label{Fig_9}
\end{figure}

We perform 10000 simulation runs of the spatiotemporal distribution outage based on both the proposed HRSRA method and the SMC method. All 936 distribution feeders of the Puerto Rico distribution network are included. The simulation period is between 0:00 UTC to 16:00 UTC on September 18, 2022, with every 10-minute simulation time step. The observed and simulated time-series power outages are shown in Fig. \ref{Fig_9}. The red curve shows the actual observed outage. The blue and orange curves show the time-series mean values of the 10000 simulation runs based on the HRSRA and the SMC method, respectively, with a 1\% to 99\% quantile range shown by the shaded area.

It can be observed from Fig. \ref{Fig_9} that the HRSRA-based simulation results precisely align with observed outages with the 1\% to 99\% quantile area of the HRSRA results covering the observation. However, the widely used SMC method exhibits inferior performance compared to the proposed HRSRA method. The significant discrepancy between the SMC result and the actual observed data is due to the SMC’s time-varying failure probability being highly sensitive to the sampling frequency. The high-frequency sampling of the time-varying failure probability leads to numerous feeder outages at an early stage, which results in substantial deviations from the actual observed outage data. In contrast, the proposed HRSRA method addresses this issue by transferring the uncertainty to the time-invariant hazard resistance of each component that is determined prior to each spatiotemporal outage simulation.

The regional spatiotemporal outage of simulation results based on the proposed HRSRA and SMC methods with a 1\% to 99\% quantile range are compared with the regional observed outages in Fig. 10. The regional comparison also validates the superiority of the proposed method. The SMC method overestimated the feeder outage in all regions.

\begin{figure}[H]
\centering
\includegraphics[width=3.5in]{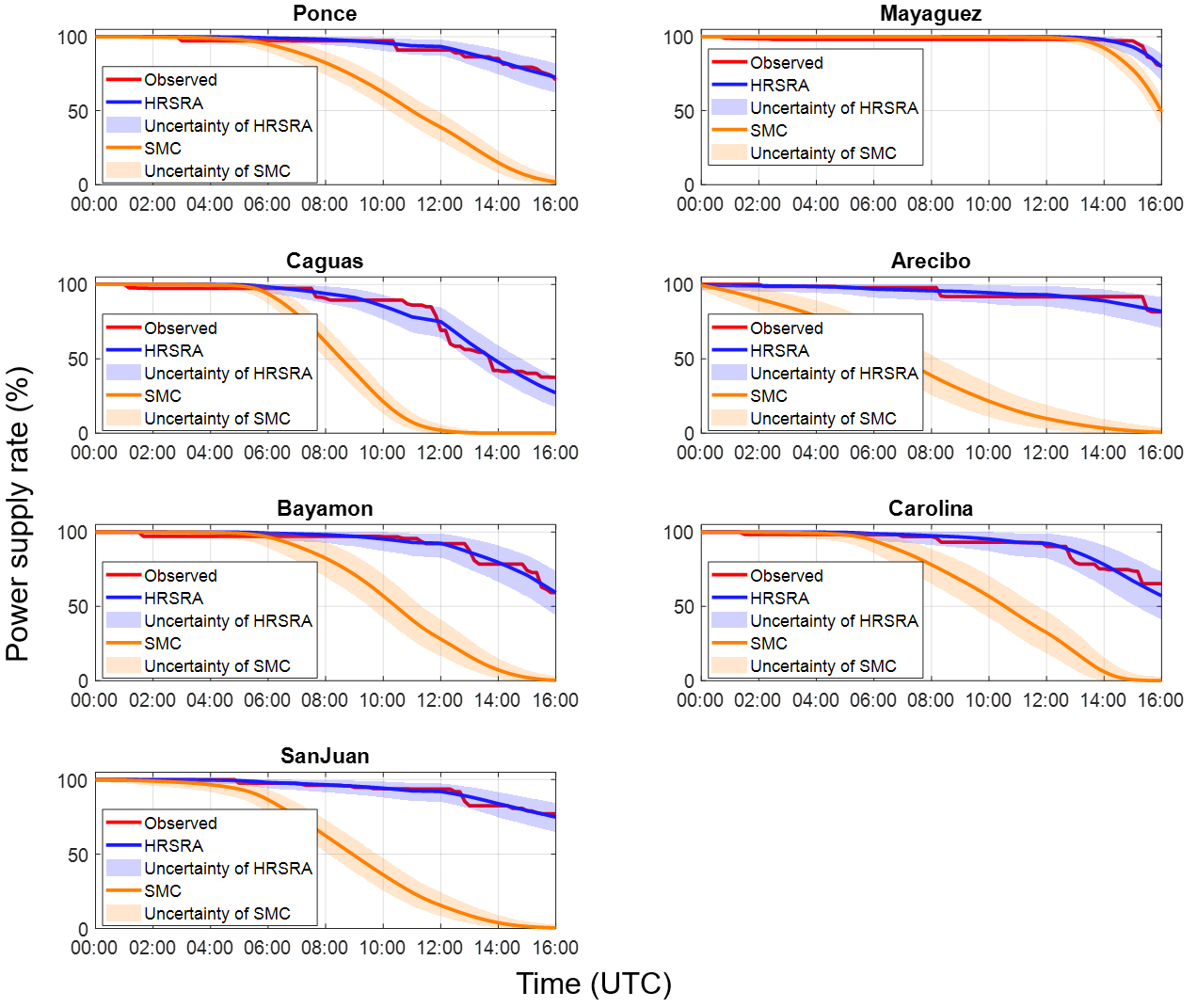}
\caption{Regional spatiotemporal outage comparison between observed and simulations based on the HRSRA and SMC methods.}
\label{Fig_10}
\end{figure}

Intuitively, an increase in the temporal resolution of simulation should improve the accuracy of a spatiotemporal risk analysis method for power system outages. To compare the precision of simulation results with actual observed data at different temporal resolutions, the following average root mean square error (RMSE) is defined:
\begin{equation}
    RMSE = \left( {\sum\limits_{i = 1}^N {\sqrt {\frac{1}{T}\sum\limits_{t = 1}^T {{{({{\bar p}_{fail}}(t) - {p_{fail,i}}(t))}^2}} } } } \right)/N
\end{equation}
\noindent where $\bar{p}_{fail}$ and $p_{fail,i}$ denote the actual observed and the i-th simulated time-series outages, respectively.

\begin{table}[!ht]
\centering
\caption{Comparison of the average $RMSE$ over 10000 simulation runs based on the HRSRA and the SMC under different temporal resolutions.}
\setlength{\tabcolsep}{13pt}
\setlength{\doublerulesep}{1pt}
\begin{tabular}{cccc}
\hline\hline
\multirow{2}*{Methods} & \multicolumn{3}{c}{$RMSE$} \\
\cline{2-4} 
 & \makecell[c]{10-min \\ resolution} &\makecell[c]{1-hour \\ resolution}  & \makecell[c]{2-hour \\ resolution}  \\
\hline
HRSRA & \textbf{1.7757} & 1.8935 & 3.3209 \\
SMC & 37.9325 & 16.1785 & 11.7423 \\
\hline
\hline
\label{Tab1}
\end{tabular}
\end{table}

Then, the average $RMSE$  of 10000 simulation runs based on the proposed and SMC methods are evaluated under different temporal resolutions such as 10-min, 1-hour, and 2-hour intervals, as shown in Table I. It can be seen from this table that with the improvement of hazard temporal resolution (from 2-hour to 10-minute resolutions), the average $RMSE$ of the proposed method decreases from 3.3209 to 1.7757 and exhibits significantly higher accuracy than the SMC results under all temporal resolutions. In addition, we can also observe the dilemma of the SMC, that is the contradiction between the simulation’s temporal resolution and the accuracy of the result. The high temporal resolution simulation, i.e., that with high sampling frequency, suffers from the ‘failure probability curse’ while the proposed HRSRA avoid this issue by converting the time-varying failure probability to the time-invariant hazard resistance of each distribution feeder in the distribution network.

\section{Conclusion}
Concerning distribution network outages during extreme weather events, this paper has presented a novel risk analysis framework for spatiotemporal power outage simulation during such persistent and temporally evolving hazards. In contrast to the current widely used Sequential Monte Carlo (SMC) simulation method, which relies on time-varying failure probability to sample component status, our proposed risk analysis method avoids the overestimation of power outage due to high-frequency sampling by applying the time-invariant hazard resistance of each component. By integrating a physics-based high-spatiotemporal-resolution hurricane wind field model and the geographic information system data of the real-world power system, the proposed spatiotemporal risk analysis method has been used to trace back the Puerto Rico power outage during Hurricane Fiona in 2022. The results from the proposed method well capture the observations. The comparison with the SMC results also validates the superiority of the proposed method under different temporal resolutions. The proposed method attains better performance at higher temporal resolutions, which indicates the simulation results are independent of sampling frequency. Thus, the proposed method offers promising application prospects under refined hazard simulations with higher spatiotemporal resolutions.

\section{Appendix A}
The parameters $\beta$ and $\lambda$ of the fragility functions for the seven regions of the lognormal CDF are calibrated by the simulated winds in mph and power outage rates in percentage during Hurricane Fiona from the Puerto Rico local power utility, as shown in Table II.
We note that the same fragility functions are applied to the distribution network outage simulations based on both the proposed HRSRA and the SMC methods.

\begin{table}[h]
    \centering
    \caption{The parameters of the fragility functions in the seven regions of Puerto Rico.}
    \setlength{\tabcolsep}{15pt}
    \setlength{\doublerulesep}{1pt}
    \begin{tabular}{ccc}
    \hline\hline
     Region    & $\lambda$ & $\beta$  \\ \hline
        Ponce & 4.7084 & 0.4379 \\
        Mayaguez & 4.4057 & 0.2061 \\
        Caguas & 4.1715 & 0.2217 \\
        Arecibo & 5.0150 & 0.8574 \\
        Bayamon & 4.4308 & 0.3012 \\
        Carolina & 4.2666 & 0.2947 \\
        San Juan & 4.4443 & 0.4226 \\
      \hline\hline    
    \end{tabular}
    \label{compare1and2}
\end{table}


\section{Biographies}
\vspace{-33pt}
 




\begin{IEEEbiographynophoto}{Luo Xu}
(S’17, M’22) received his B.S. and Ph.D. degrees both in electrical engineering from Wuhan University, China in 2017 and Tsinghua University, China in 2022, respectively. He is a Postdoctoral Researcher at the Department of Civil and Environmental Engineering, Princeton University. His current research interests include power system resilience under climate risks, resilience, and stability analysis of digital power systems. He serves as the officer of the IEEE PES Energy Internet Coordinating Committee (EICC), the Secretary of CIGRE Working Group D2.56 on “Interdependence and Security of Cyber-Physical Power System” and a member of the IEEE Task Force on “Cyber-Physical Interdependence for Power System Operation and Control”. He is a Young Editorial Board Member of \textit{Applied Energy}, and a reviewer for 13 journals including \textit{IEEE Transactions on Smart Grid} and \textit{IEEE Transactions on Power Systems}. He was a recipient of CIGRE Thesis Award 2023, Best Research Award of IEEE PES PhD Dissertation Challenge in 2023, the IET Premium Awards in 2019 and the Best Conference Paper Award from the IEEE Conference on Energy Internet and Energy System Integration in 2017, and the Outstanding Reviewer for \textit{IEEE Transactions on Power Systems} in 2020.
\end{IEEEbiographynophoto}

\begin{IEEEbiographynophoto}{Ning Lin}
is a Professor of Civil and Environmental Engineering at Princeton University, where she has affiliate appointments with Princeton School for Public and International Affairs, Andlinger Center for Energy and Environment, and High Meadows Environmental Institute. Lin’s research areas include Natural Hazards and Risk Analysis, Wind Engineering, Coastal Engineering, and Climate Change Impact and Adaptation. Her current primary focus is hurricane risk analysis. She integrates science, engineering, and policy to study hurricane-related weather extremes (strong winds, heavy rainfall, and storm surges, and compounding sea level rise and heatwaves), how they change with changing climate, and how their impact on society can be better mitigated. She is a recipient of CAREER award from National Science Foundation (NSF), Natural Hazards Early Career Award and Global Environmental Change Early Career Award from American Geophysical Union (AGU), and Huber Research Prize from American Society of Civil Engineers (ASCE). Lin has been the lead PI or Co-PI for several large NSF projects, including Interdisciplinary Research in Hazards and Disasters (Hazards SEES), Prediction of and Resilience against Extreme Events (PREEVENTS), and Coastlines and People Hubs for Research and Broadening Participation (CoPe). Lin received her Ph.D. in Civil and Environmental Engineering from Princeton University in 2010. She also received a certificate in Science, Technology and Environmental Policy in 2010 from Princeton. Before rejoining Princeton as an assistant professor in 2012, she conducted research in the Department of Earth, Atmospheric and Planetary Sciences at MIT as a NOAA Climate and Global Change Postdoctoral Fellow.
\end{IEEEbiographynophoto}

\begin{IEEEbiographynophoto}{Dazhi Xi}
is an Associate Research Scholar at Princeton University. He obtained his Ph.D. degree from the Department of Civil and Environmental Engineering from Princeton University and his Bachelor of Science degree from the School of Atmospheric Sciences at Nanjing University. His research focuses on the physical science of climate change, statistical climate downscaling, and hazard assessment of tropical cyclones and heatwaves.
\end{IEEEbiographynophoto}

\begin{IEEEbiographynophoto}{Kairui Feng}
 is currently a Postdoctoral Research Associate at the Department of Civil and Environmental Engineering, Princeton University, New Jersey, USA. He received a Ph.D. degree in Civil and Environmental Engineering from Princeton University and bachelor's degrees in Civil Engineering and Mathematics from Tsinghua University, China. His research interest includes infrastructure system modeling/optimization and climate change using data-driven and numerical approaches.
\end{IEEEbiographynophoto}

\begin{IEEEbiographynophoto}{H. Vincent Poor}
(S’72, M’77, SM’82, F’87) received the Ph.D. degree in EECS from Princeton University in 1977.  From 1977 until 1990, he was on the faculty of the University of Illinois at Urbana-Champaign. Since 1990 he has been on the faculty at Princeton, where he is currently the Michael Henry Strater University Professor. During 2006 to 2016, he served as the dean of Princeton’s School of Engineering and Applied Science. He has also held visiting appointments at several other universities, including most recently at Berkeley and Cambridge. His research interests are in the areas of information theory, machine learning and network science, and their applications in wireless networks, energy systems and related fields. Among his publications in these areas is the book \textit{Advanced Data Analytics for Power Systems}. (Cambridge University Press, 2021). Dr. Poor is a member of the National Academy of Engineering and the National Academy of Sciences and is a foreign member of the Chinese Academy of Sciences, the Royal Society, and other national and international academies. He received the IEEE Alexander Graham Bell Medal in 2017.
\end{IEEEbiographynophoto}

\vfill

\end{document}